\documentclass[preprint,showpacs,preprintnumbers,amsmath,amssymb,nofootinbib,showkeys,aps]{revtex4}
\pdfoutput=1
\usepackage{epsfig}

\usepackage{graphicx}
\usepackage{dcolumn}
\usepackage{bm}
\usepackage{amsmath}
\usepackage{amssymb}
\usepackage{latexsym}
\usepackage{color}
\usepackage{hyperref}
\usepackage{mathrsfs}
\usepackage{float}

\newcommand{\be}{\begin{equation}}
\newcommand{\ee}{\end{equation}}
\newcommand{\bea}{\begin{eqnarray}}
\newcommand{\eea}{\end{eqnarray}}
\newcommand{\texpdf}{\texorpdfstring}

\begin{document}


\title{Can decaying vacuum solve the $H_0$ Tension?}

\author{L. S. Brito$^{1}$} \email{lucas.s.brito@unesp.br}
\author{J. F. Jesus$^{1,2}$}\email{jf.jesus@unesp.br}
\author{A. A. Escobal$^{1,3,4}$}\email{anderson.aescobal@gmail.com}
\author{S. H. Pereira$^{1}$}\email{s.pereira@unesp.br}

\affiliation{$^1$Universidade Estadual Paulista (UNESP), Faculdade de Engenharia e Ci\^encias, Departamento de F\'isica - Av. Dr. Ariberto Pereira da Cunha 333, 12516-410, Guaratinguet\'a, SP, Brazil
\\$^2$Universidade Estadual Paulista (UNESP), Instituto de Ciências e Engenharia, Departamento de Ci\^encias e Tecnologia - R. Geraldo Alckmin, 519, 18409-010, Itapeva, SP, Brazil\\
$^3$Department of Astronomy, University of Science and Technology of China, Hefei, Anhui 230026, China\\
$^4$IAG, Universidade de S\~ao Paulo, 05508-900 S\~ao Paulo, SP, Brazil.
}


\def\zt{\mbox{$z_t$}}

\begin{abstract}
In the present work we analyze two different models of interaction between dark energy and dark matter, also known as vacuum decay models or $\Lambda(t)$CDM models. In both models, when the $H_0$ parameter is constrained by {high-redshift data as the Planck distance priors (CMB) and transversal BAO}, its value is compatible with a higher value of $H_0$, in agreement with low-redshift data, as Pantheon+\&SH0ES (PS) and Cosmic Chronometers (CC). For one of the models, only a mild $\sim2.1\sigma$ tension is found for $\Omega_m$, which at least is an alleviation to the $\gtrsim5\sigma$ $H_0$ tension {in the context of standard $\Lambda$CDM model}. We find $H_0=71.63\pm0.60$ km/s/Mpc for one model and $H_0=71.67\pm0.60$ km/s/Mpc for the other one, by combining CC+PS+BAO+CMB. We also find the decay parameter to be $\epsilon=0.0162^{+0.0029}_{-0.0026}$ for one model and $\epsilon=0.0209\pm0.0036$ for the other one. All constraints are at 68\% c.l. From these analyses, a noninteracting model is excluded at least at $5.8\sigma$ c.l. {We shall emphasize, however, that this result comes from an analysis that involves an approximated treatment to CMB, the Planck distance priors, as mentioned above.} Furthermore, while we have included the SH0ES constraints on the present analysis, other analyses in the literature not including SH0ES do not obtain this level of significance for an interacting model. This shows that these types of models are promising in solving or at least alleviating the $H_0$ tension problem. Our analysis also shows a preference for the decay of vacuum into dark matter, in agreement to thermodynamic analyses.

\end{abstract}

\maketitle



\section{Introduction} 

The $\Lambda$CDM model (formed by a {Cold Dark Matter} component plus a {cosmological constant} $\Lambda$) is currently the most widely accepted cosmological model to describe the evolution of the Universe, due to its ability to accurately describe key observational data, such as the cosmic microwave background (CMB) power spectrum and the accelerated expansion of the Universe. However, despite its success, the $\Lambda$CDM model faces several challenges, both theoretical and observational. Among these are the cosmological constant problem or Dark Energy (DE) problem, the nature of Dark Matter (DM), the coincidence problem and more recently the Hubble tension.

The cosmological constant problem arises from the discrepancy between theoretical predictions of vacuum energy from quantum field theory and the observed value of the vacuum energy density. Quantum Field Theory in curved spacetime predicts a vacuum energy density that is approximately 120 orders of magnitude larger than what is inferred from astronomical observations. This vast difference between the theoretical prediction and the observationally inferred value remains one of the greatest unsolved problems in modern Cosmology \cite{Weinberg:1988cp,Perivolaropoulos2022}. Another profound mystery in modern Cosmology is the nature of DM. Although it constitutes approximately 27\% of the total mass-energy content of the Universe, dark matter does not interact with electromagnetic radiation, making it invisible to current detection methods. Its presence is inferred primarily through its gravitational effects, such as the rotation curves of galaxies and gravitational lensing. Despite extensive efforts to directly detect dark matter particles, their exact nature remains elusive. The coincidence problem, on the other hand, refers to the seemingly peculiar fact that the current densities associated with the DE or vacuum energy and DM are of the same order of magnitude, despite having evolved very differently throughout the history of the Universe. In the early Universe, the density of matter was vastly greater than that of dark energy. As the Universe expanded, dark matter density diluted, while dark energy remained constant. The challenge is to explain why, out of all possible epochs, we observe the Universe at a time when these densities appear so closely aligned \cite{Velten:2014nra}.

In this work, however, we focus on another significant cosmological issue: the Hubble tension \cite{H0TensionReview}. Early estimates of the Hubble constant ($H_0$), which describes the current expansion rate of the Universe, varied considerably. During the 1960s and 1970s, measurements ranged from 50 km/s/Mpc to 100 km/s/Mpc, mainly due to the limitations in observational techniques at that time. In the 1990s, the Hubble Space Telescope provided more precise estimates of $H_0$, leading to a period of significant improvement in measurements. The SH0ES (Supernovae, $H_0$ for the Dark Energy Equation of State) Program further refined the value of the $H_0$ through detailed analyses of parallax measurements, Cepheid variable stars, gravitational lensing, and Type Ia supernovae (SNe Ia). The SH0ES team determined $H_0$ to be $74.2\pm3.6$ km/s/Mpc \cite{Riess:2009pu}, with subsequent studies improving the precision to $73.17\pm0.86$ km/s/Mpc \cite{BreuvalEtAl24}. This value will be referred to as the SH0ES $H_0$. In contrast, the Planck satellite mission, operated by the European Space Agency (ESA), used a different approach to measure the expansion rate of the Universe. By analyzing data from the CMB in combination with baryon acoustic oscillation (BAO) data, the Planck team derived a lower value of $H_0=67.4\pm 0.5$ km/s/Mpc \cite{Planck18}, which we refer to as the Planck $H_0$.

The discrepancy between the SH0ES and Planck measurements of $H_0$ has sparked a significant debate within the cosmology community, as both values are derived from highly robust and independent datasets. The SH0ES estimate is based on local or low-redshift observations, while the Planck value stems from early or high-redshift Universe data, under the assumption of the $\Lambda$CDM model. This discrepancy has now reached a 5.8$\sigma$ significance level, posing a serious challenge to the $\Lambda$CDM framework. Since systematic errors have not been able to account for this tension, various theoretical models have been proposed to resolve the issue. These include modifications such as models with additional degrees of freedom in the dark energy sector, and other alternative cosmological proposals \cite{H0TensionReview}. 

In the present paper we are interested to investigate the Hubble tension in a special class of DE models, namely models with a time-varying $\Lambda$\cite{Ahmet2018,Vishwakarma2001,Overduin1998,Azri2017,Azri2012,Szydlowski2015,Bruni2022,Papagian2020,Benetti2019,Benetti2021,Macedo:2023zrd}. In such models, some \textit{ad hoc} time dependence for $\Lambda(t)$ is assumed or can also be derived from Quantum Mechanical arguments or by geometrical motivations. {There is also a class of models which can be derived from fundamental quantum field theories calculations in curved spacetime, known as running vacuum models (RVM). The theoretical development of RVM within the framework of quantum field theory has been addressed in several works, including \cite{Moreno-Pulido:2020anb,Moreno-Pulido:2022phq,Moreno-Pulido:2023ryo}. While the general form of RVM has been known for some time from semi-qualitative renormalization group analyses (see the early review \cite{Sola:2013gha} and references therein), only more recently have explicit quantum field theory computations provided rigorous derivations of these formulas, as discussed in the modern review \cite{SolaPeracaula:2022hpd}. Relatively recent phenomenological analyses of these models were carried out in the references \cite{Sola:2017znb,SolaPeracaula:2017esw,SolaPeracaula:2021gxi,SolaPeracaula:2023swx,Sola:2016jky,Sola:2015wwa}.

In the context of running cosmological constant models, one of the first confrontations with observational data, specifically SNe Ia data, was presented in the works 
\cite{Espana-Bonet:2003qjh,Shapiro:2003ui}. In these studies, semiclassical models were developed in which the running cosmological constant evolves smoothly, without fine-tuning, as a consequence of generic quantum effects near the Planck scale. It was demonstrated that, due to the decoupling phenomenon, low-energy effects are irrelevant for the running of the cosmological constant. Consequently, the approximate coincidence between the density parameters of dark matter and dark energy is not restricted to any specific epoch in the history of the Universe. Subsequently, similar models of interacting dark energy and dark matter were explored in detail in references \cite{SolaPeracaula:2017esw} and \cite{Sola:2017znb}. These studies provided analytical solutions and extensive numerical analyses, incorporating a wide range of observational data, including SNe Ia, BAO, $H(z)$, LSS, and CMB measurements. In the first reference, the running vacuum model was identified as the most theoretically motivated and statistically favoured, with a confidence level of $3\sigma$. These results were further supported by the computation of the Akaike and Bayesian information criteria. In the second reference, the Hubble tension was analyzed in the context of vacuum dynamics, showing that the values of $H_0$ inferred from CMB observations are more strongly favoured than those obtained from local measurements. {In \cite{Sola:2016jky,Sola:2015wwa} it was shown a significantly better agreement of RVM as compared to the concordance model, including structure formation data.}

{Here we analyze two different models of interaction between dark energy and dark matter. The first model has been extensively studied in various contexts; however, in this work, we focus specifically on the $H_0$ tension. For the second proposed model, to the best of our knowledge, it is being investigated for the first time.}

The paper is organized as follows. We describe the Friedmann equations with a time-dependent $\Lambda(t)$-term in Section \ref{sec: Friedmann eqs}, where we also obtain analytical solutions for $H(z)$ in different classes of models. In Section \ref{sec: analysis}, we constrain the parameters of the models using {CMB from Planck, Pantheon+\&SH0ES SNe Ia, Cosmic Chronometers (CC) and BAO and also compare the decaying vacuum models with standard $\Lambda$CDM. In the analysis, we consider separately the constraints from low redshift and high redshift data.} Finally, we present our conclusions and final remarks in Section \ref{sec: conclusion}.

\section{\label{sec: Friedmann eqs} Cosmological equations for a varying \texpdf{\boldmath{$\Lambda$}}{L} term}
From the cosmological principle, which assures that the spatial distribution of matter in the Universe is uniformly homogeneous and isotropic when viewed on large scale, the Einstein field equations leads to the so-called Friedmann equations, given by:
\begin{align}
H^2 &= \frac{8\pi G \rho_T}{3} - \frac{k}{a^2},\label{fried0}\\
\frac{\ddot{a}}{a} &= -\frac{4\pi G}{3}(\rho_T + 3p_T),
\label{fried}
\end{align}
where $\rho_T$ is the total energy density of the Universe, $p_T$ is the total pressure and $k$ is the spatial curvature, which we assume to be zero from now on. Taking into account different contributions to $\rho_T$ and $p_T$, they are given by:
\begin{align}
\rho_{T} &= \rho_{M}+\rho_{\Lambda} + \rho_r,\\
p_T &=p_\Lambda+p_r=-\rho_\Lambda+\frac{1}{3}\rho_r
\end{align}
where $\rho_M = \rho_d + \rho_b$ corresponds to the sum of dark matter energy density ($\rho_d$) plus baryons energy density ($\rho_b$), which are both pressureless ($p_M=0$),  $\rho_\Lambda$ corresponds to the time-varying $\Lambda(t)$-term, (with $p_\Lambda = - \rho_\Lambda$) and $\rho_r$ corresponds to radiation, (with $p_r = \frac{1}{3} \rho_r$). The continuity equations may be written as:
\begin{align}
\dot{\rho}_{b} + 3H{\rho_{b}} &= 0,\label{rhob}\\
\dot{\rho}_{r} + 4H{\rho_{r}} &= 0,\label{rhor}\\
\dot{\rho}_d + 3H{\rho_d} &= Q,\\
\dot{\rho}_{\Lambda} &= -Q,\label{rhoL}
\end{align}
where $Q$ is the interaction term between pressureless matter and vacuum. Eqs. \eqref{rhob} and \eqref{rhor} promptly yield the standard result $\rho_b=\rho_{b0}a^{-3}$ and $\rho_r=\rho_{r0}a^{-4}$. In order to solve for $\rho_d$ and $\rho_\Lambda$, however, we need the explicity form of the interaction term $Q$. Since we knows that baryons and dark matter share the same equation of state (EOS), one might also write:
\be
\dot{\rho}_M + 3H{\rho_M} = Q
\label{rhoM}
\ee
for the total pressureless matter. 

In the present work we will study two possible interaction terms, namely $Q=3\epsilon H\rho_\Lambda$ (Model I) and $Q=3\epsilon aH\rho_\Lambda$ (Model II). Thus, one could summarize both interaction terms in one equation:
\be
Q=3\epsilon a^nH\rho_\Lambda
\label{Qgeral}
\ee
where $n=0$ or $n=1$ for the Models I and II, respectively. Model I is interesting for dimensional reasons, as the interaction term has dimensions $\propto H\rho$. Model II is interesting because it corresponds to an exponential decay of $\Lambda$ with respect to the scale factor, as we shall see below. In the following subsections, we study each of these cases separately.

\subsection{Model I: \texpdf{\boldmath{$Q=3\epsilon H\rho_\Lambda$}}{Q=3*epsilon*H*rho Lambda} class of models}\label{sec: Hrho}
This model has already been studied by \cite{SolaPeracaula:2017esw,PavonZimdahl05,PavonWang09,PereiraJesus09}, in the context of a dark energy with a constant equation of state. In this case, by replacing $n=0$ in Eq. \eqref{Qgeral}, Eq. \eqref{rhoL} can be written as:
\be
\dot{\rho}_{\Lambda} = -3\epsilon H\rho_\Lambda
\ee
By changing to scale factor derivatives by using the relation $\frac{d}{dt}=\dot{a}\frac{d}{da}=aH\frac{d}{da}$, it can be written as:
\be
\rho_\Lambda'(a)=-\frac{3\epsilon\rho_\Lambda(a)}{a}
\ee
where $'$ means $\frac{d}{da}$. This equation can easily be solved to:
\be
\rho_\Lambda=\rho_{\Lambda0}a^{-3\epsilon}
\label{rhoLn0}
\ee
where $\rho_{\Lambda0}$ is the current value of $\Lambda$ density. This coincides with the so-called Generalized Chen-Wu model, where $\Lambda\propto a^{-n}$ \cite{JohnJoseph99,Chen1990}. By replacing Eq. \eqref{rhoLn0} in Eq. \eqref{rhoM}, and changing to scale factor derivatives, we find:
\be
\rho_M'(a)=-\frac{3\rho_M}{a}+3\epsilon\rho_{\Lambda0}a^{-1-3\epsilon}\,.
\ee
This equation can be solved to:
\be
\rho_M(a)=\left(\rho_{M0}-\frac{\epsilon\rho_{\Lambda0}}{1-\epsilon}\right)a^{-3}+\frac{\epsilon\rho_{\Lambda0}}{1-\epsilon}a^{-3\epsilon}\,.
\label{rhoMn0}
\ee

By replacing \eqref{rhoLn0} and \eqref{rhoMn0} in Friedmann equation \eqref{fried0} and changing to redshift $z$, we find
\be
H^2=H_0^2\left[\left(\Omega_{M}-\frac{\epsilon\Omega_{\Lambda}}{1-\epsilon}\right)(1+z)^{3}+\frac{\Omega_{\Lambda}(1+z)^{3\epsilon}}{1-\epsilon}+\Omega_{r}(1+z)^4\right]
\ee
where $\Omega_i\equiv\frac{8\pi G\rho_{i0}}{3H_0^2}$ are the current density parameters and the normalization condition reads:
\be
\Omega_M+\Omega_\Lambda+\Omega_r=1\,.
\ee

\subsection{Model II: \texpdf{\boldmath{$Q=3\epsilon aH\rho_\Lambda$}}{Q=3*epsilon*conformal H*rho Lambda} class of models}\label{sec: Hconfrho}
In this case, $n=1$ in Eq. \eqref{Qgeral} and Eq. \eqref{rhoL} can be solved to:
\be
\rho_\Lambda=\rho_{\Lambda*}e^{-3\epsilon a}
\label{rhoLExp}
\ee
where
\be
\rho_{\Lambda*}\equiv\rho_{\Lambda0}e^{3\epsilon}
\label{rhoLExp*}
\ee

As one may see, Eq. \eqref{rhoLExp} corresponds to an exponential decay of $\Lambda$ with respect to the scale factor. This model has already been proposed by Rajeev \cite{Rajeev83}. Although being proposed decades ago, it has never been tested against observations, as we shall test below.

Replacing \eqref{rhoLExp} and \eqref{rhoLExp*} into Eq. \eqref{rhoM} and solving, we find:
\be
\rho_M=\frac{\rho_{M0}}{a^3}+\frac{\rho_{\Lambda0}}{9\epsilon^3a^3}\left[P(\epsilon)-P(\epsilon a)e^{3\epsilon(1-a)}\right]\,,
\label{rhoMLExp}
\ee
where the cubic polynomial $P(x)$ is defined as:
\be
P(x)\equiv9x^3+9x^2+6x+2\,.
\ee

By replacing \eqref{rhoLExp} and \eqref{rhoMLExp} together with radiation term into Friedmann equation \eqref{fried} we find:
\be
H^2=\frac{8\pi G}{3}\left\{\frac{\rho_{M0}}{a^3}+\frac{\rho_{\Lambda0}}{9\epsilon^3a^3}\left[P(\epsilon)-P(\epsilon a)e^{3\epsilon(1-a)}\right]+\rho_{\Lambda0}e^{3\epsilon(1-a)}+\rho_{r0}a^{-4}\right\}\,
\ee
or:
\be
H^2=\frac{8\pi G}{3}\left\{\frac{\rho_{M0}}{a^3}+\frac{\rho_{\Lambda0}}{9\epsilon^3a^3}\left[P(\epsilon)-Q(\epsilon a)e^{3\epsilon(1-a)}\right]+\rho_{r0}a^{-4}\right\}\,,
\label{H2aLExp}
\ee
where $Q(x)$ is the quadratic polynomial:
\be
Q(x)\equiv P(x)-9x^3=9x^2+6x+2\,.
\ee
Dividing \eqref{H2aLExp} by $H_0^2$, we find:
\be
E^2\equiv\frac{H^2}{H_0^2}=\frac{\Omega_{M}}{a^3}+\frac{\Omega_{\Lambda}}{9\epsilon^3a^3}\left[P(\epsilon)-Q(\epsilon a)e^{3\epsilon(1-a)}\right]+\Omega_{r}a^{-4}\,,
\ee
where $\Omega_i\equiv\frac{8\pi G\rho_{i0}}{3H_0^2}$ are the current density parameters and the normalization condition reads:
\be
\Omega_M+\Omega_\Lambda+\Omega_r=1\,.
\ee
In terms of $z$, we have:
\be
E(z)^2=\Omega_{M}(1+z)^3+\frac{\Omega_{\Lambda}(1+z)^3}{9\epsilon^3}\left[P(\epsilon)-Q\left(\frac{\epsilon}{1+z}\right)e^{\frac{3\epsilon z}{1+z}}\right]+\Omega_{r}(1+z)^4\,.
\ee

\section{\label{sec: analysis} Analysis and Results}


In Ref. \cite{BreuvalEtAl24}, the SH0ES collaboration were able to improve the cosmic distance ladder by calibrating the period-luminosity relation with photometric measurements of 88 Cepheid variables in the core of the Small Magellanic Cloud. With this analysis, they were able to increase the foundation of the cosmic distance ladder from three to four calibrating galaxies, thereby obtaining a combined result of $H_0=73.17\pm0.86$ km/s/Mpc. With this result, as explained by \cite{BreuvalEtAl24}, the local measurement of $H_0$ based on Cepheids and Type Ia Supernovae shows a 5.8$\sigma$ tension with the value inferred from CMB in the context of a $\Lambda$CDM cosmology, $H_0=67.4\pm0.5$ km/s/Mpc \cite{Planck18}. {As we intend to add the most recent SNe Ia compilation, Pantheon+, we have chosen to use the Pantheon+\&SH0ES (PS) data as representing low reshift data, including constraints over $H_0$. In fact, \cite{pantheon+} have found $H_0=73.6\pm1.1$ km/s/Mpc in the context of flat $\Lambda$CDM model, which already indicated a 5.1$\sigma$ tension.}

{As another relevant low-redshift dataset, we chose to work with Cosmic Chronometers (CC) data, which is interesting because it is independent of cosmological models, relying only on Astrophysical assumptions. We have chosen as CC sample, the recent sample from Moresco \textit{et al.}, consisting of 32 $H(z)$ data with estimated statistical and systematic uncertainties \cite{MorescoEtAl22}.}

Here, instead of working with the full CMB power spectrum, we work simply with the so-called distance priors from Planck \cite{ChenEtAl19}, which consist on three cosmological quantities obtained from the Planck 2018 Monte Carlo chains and, as shown by \cite{ChenEtAl19}, result in approximately the same constraints from the full spectrum, in the context of flat $\Lambda$CDM, O$\Lambda$CDM and flat XCDM.

{In order to complement the high-redshift data, we chose to work with Baryon Acoustic Oscillations (BAO) constraints, in the form of the transversal BAO data \cite{NunesEtAl20}. We chose this dataset because it is claimed to be less model-dependent, which is crucial for BAO analyses, since most of BAO constraints tend to be model-dependent, introducing biases towards standard models.}

In all analyses we have assumed a flat prior over the free parameters: $\Omega_m\in[0,1]$, $H_0\in[10,100]$ km/s/Mpc, $\Omega_b\in[0.02,0.07]$ and $\epsilon\in[-0.5,0.5]$. It is important to mention that we have also allowed for a negative range for $\epsilon$. However, as we shall see below, this range will not be favoured when combining the {the low-redshift data and high-redshift data. In order to sample the posteriors, we have used the software {\sffamily emcee}\footnote{\url{https://emcee.readthedocs.io/}} \cite{emcee} and in order to plot the results, we have used the software {\sffamily getdist}\footnote{\url{https://getdist.readthedocs.io/}} \cite{getdist}.}



\begin{figure}[t]
    \centering
    \includegraphics[width=0.9\textwidth]{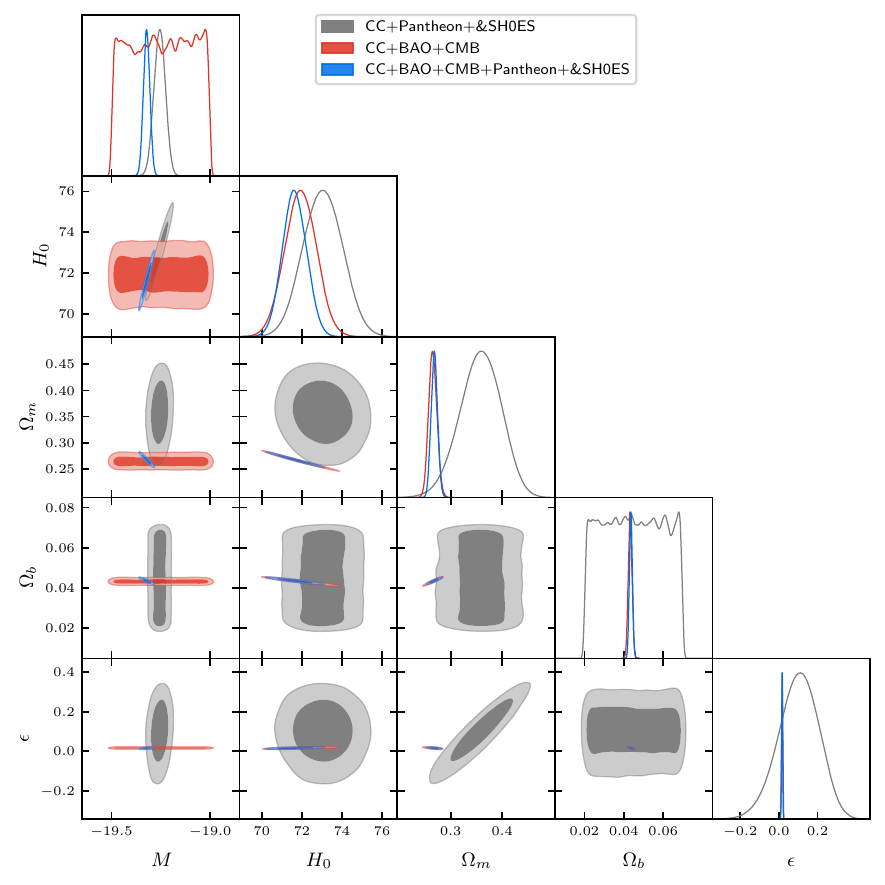}
    \caption{Triangular plot of parameters, with data from Planck 2018 distance priors {(CMB), Pantheon+\&SH0ES, Cosmic Chronometers (CC) and transversal BAO} for Model I (with $Q= 3\epsilon H\rho_\Lambda$). The contours correspond to 68\% and 95\% c.l.}\label{Tplot1}
\end{figure}

\subsection{Model I (\texorpdfstring{$Q=3\epsilon H\rho_\Lambda$}{Q=3*epsilon*H*rhoLambda}) results}

In Fig. \ref{Tplot1} it is shown the results for Model I ($Q=3\epsilon H\rho_\Lambda$). {We chose to work with the combinations CC+Pantheon+\&SH0ES as low redshift constraints and CC+BAO+CMB as high-redshift constraint. Although CC can not be considered as high redshift constraint, it helps with the convergence of the MC chains, even if it provides weak constraints. As can be seen from this Figure, the combination CC+BAO+CMB strongly constrain $H_0$, $\Omega_m$, $\Omega_b$ and $\epsilon$. It can be seen that $\epsilon>0$ favours a higher value for $H_0$. Although most of the constraints are compatible for low and high redshift data, one may notice a slight tension for $\Omega_m$: low-redhift data favour a higher value of $\Omega_m$, while high-redshift data favour a lower value. In order to better quantify this tension, let us have a look at Table \ref{tab1}.}

\begin{table}[ht]
    \centering
    \begin{tabular} {l | c | c | c | c}

Dataset & $H_0$(km/s/Mpc) & $\Omega_m$ & $\Omega_b$ & $\epsilon$\\  
\hline  
CC+PS & $73.05\pm 0.97$ & $0.358^{+0.042}_{-0.038}$ & $0.045\pm 0.014$ & $0.101^{+0.11}_{-0.097}$\\
CC+BAO+CMB & $71.93\pm 0.79$ & $0.2650^{+0.0075}_{-0.0084}$ & $0.04322\pm 0.00094$ & $0.0170^{+0.0031}_{-0.0027}$\\
CC+PS+BAO+CMB &$71.63\pm 0.60$&$0.2682\pm 0.0062$ &$0.04361\pm 0.00075$ & $0.0162^{+0.0029}_{-0.0026}$\\
\hline
\end{tabular}
    \caption{{Numerical results for Model I. Mean values of the free parameters from the various combinations of datasets, together with the $1\sigma$ c.l. intervals.}}
    \label{tab1}
\end{table}

{As can be seen from this Table, all parameters are compatible, except for $\Omega_m$, where a small tension can be found of 2.2$\sigma$. Although being a significant tension, it is lower than the $H_0$ tension within $\Lambda$CDM of 5.1$\sigma$. We may say that the tension is alleviated within Model I. Fig. \ref{Tplot2ModI} shows the combined constraints for Model I. A positive value for $\epsilon$ is favoured from this analysis.}


\begin{figure}[t]
    \centering
    \includegraphics[width=0.9\textwidth]{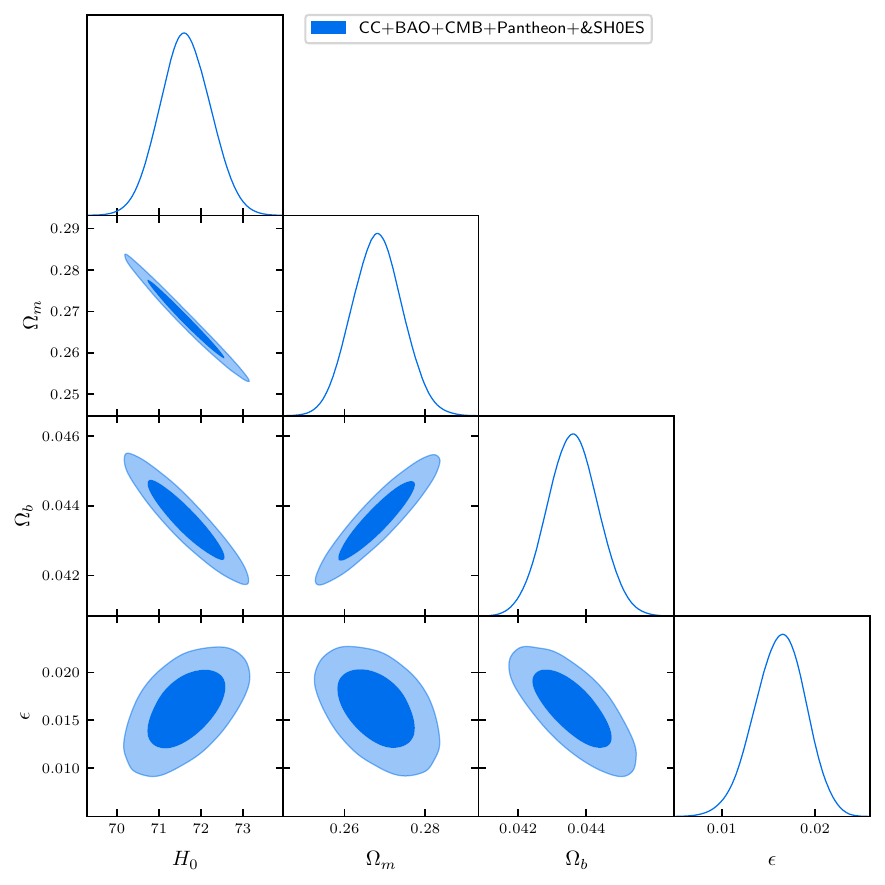}
    \caption{{Triangular plot of parameters for the joint analysis of CC+BAO+CMB+Pantheon+\&SH0ES for Model I (with $Q= 3\epsilon H\rho_\Lambda$). The contours correspond to 68\% and 95\% c.l.}}\label{Tplot2ModI}
\end{figure}

\begin{figure}[t]
    \centering
    \includegraphics[width=0.9\textwidth]{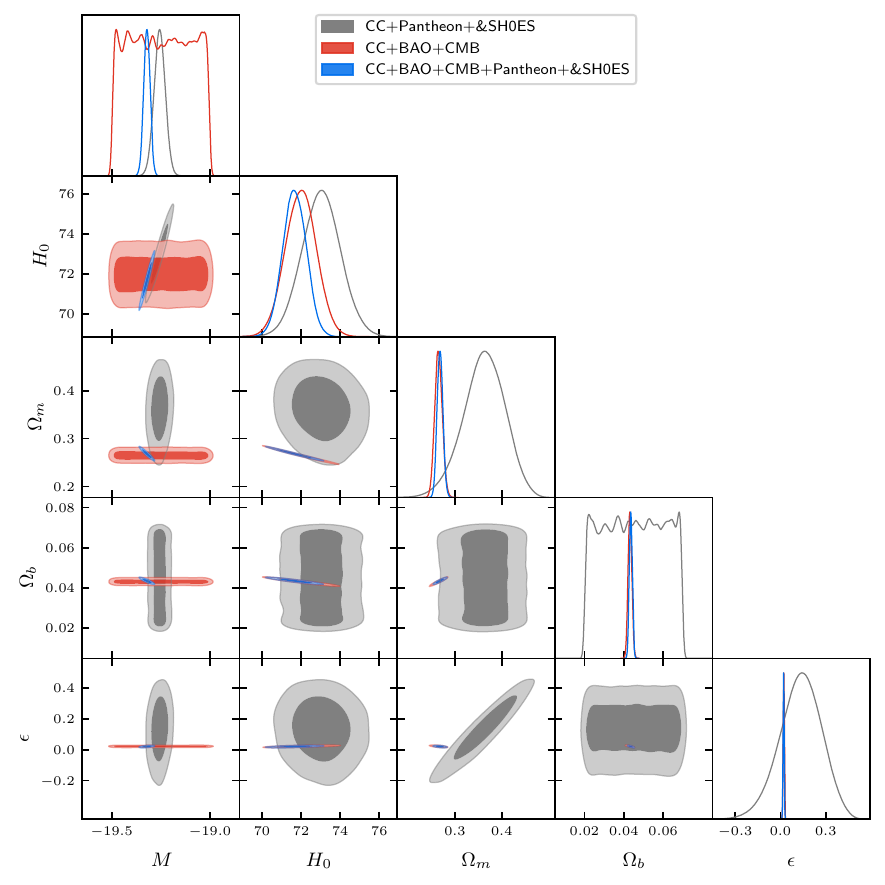}
    \caption{
{Triangular plot of parameters, with data from Planck 2018 distance priors {(CMB), Pantheon+\&SH0ES, Cosmic Chronometers (CC) and transversal BAO} for Model II (with $Q= 3\epsilon aH\rho_\Lambda$). The contours correspond to 68\% and 95\% c.l.}}\label{Tplot2}
\end{figure}

\newpage
\subsection{Model II (\texorpdfstring{$Q=3\epsilon aH\rho_\Lambda$}{Q=3*epsilon*a*H*rhoLambda}) results}

In Fig. \ref{Tplot2}, we have the constraints for Model II ($Q=3\epsilon aH\rho_\Lambda$). {In general, the constraints are quite similar to Model I, but the high-redshift constraint over $\Omega_m$ seems to be slightly more consistent with the low-redshift constraint.}

\begin{table}[ht]
    \centering
    \begin{tabular} {l | c | c | c | c}

Dataset & $H_0$(km/s/Mpc) & $\Omega_m$ & $\Omega_b$ & $\epsilon$\\
\hline
CC+PS & $73.04\pm0.98$ & $0.362^{+0.048}_{-0.041}$ & $0.045\pm 0.015$ & $0.13^{+0.15}_{-0.13}$\\
CC+BAO+CMB & $71.99\pm 0.79$ & $0.2650\pm0.0080$ & $0.04316\pm0.00095$ & $0.0221^{+0.0040}_{-0.0035}$\\
CC+PS+BAO+CMB & $71.67\pm 0.60$ & $0.2683\pm0.0061$ & $0.04357\pm0.00075$ & $0.0209\pm0.0036$\\
\hline
\end{tabular}
    \caption{{Numerical results for Model II. Mean values of the free parameters from the various combinations of datasets, together with the $1\sigma$ c.l. intervals.}}
    \label{tab2}
\end{table}

\newpage

{As can be seen on this table, the values for all parameters are compatible between low-redshift and high-redshift data, except for $\Omega_m$, where a 2.1$\sigma$ tension was found. Again, we may say that the $H_0$ tension is transferred and alleviated to $\Omega_m$.}

\begin{figure}[t]
    \centering
    \includegraphics[width=0.9\textwidth]{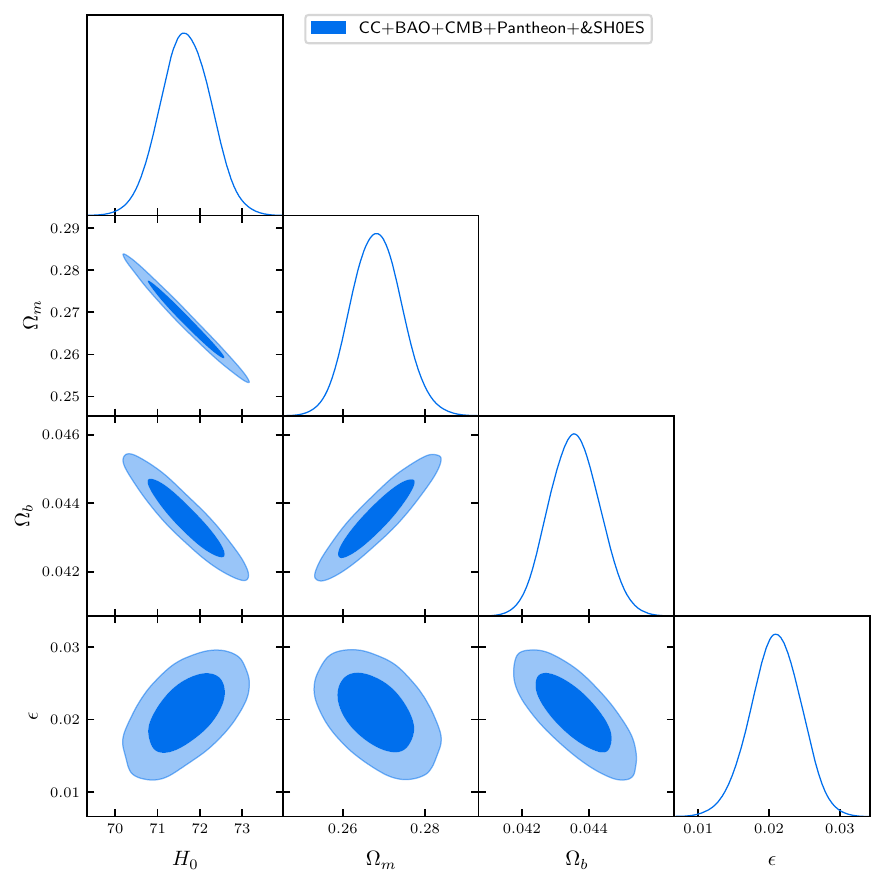}
    \caption{{Triangular plot of parameters for the joint analysis of CC+BAO+CMB+Pantheon+\&SH0ES for Model II (with $Q= 3\epsilon aH\rho_\Lambda$). The contours correspond to 68\% and 95\% c.l.}}\label{Tplot2ModII}
\end{figure}


{Note that the value for $\epsilon=0.0162^{+0.0029}_{-0.0026}$ for Model I excludes a noninteracting model with $\sim6.2\sigma$ c.l. The situation is quite similar for Model II, where $\epsilon=0.0209\pm0.0036$ excludes a noninteracting model at $\sim5.8\sigma$.}

There are some analyses available in the literature concerning Model I. Bernui \textit{et al.} \cite{BernuiEtAl23} analyse a model similar to Model I, as they assume a dark energy with EOS $w=-0.9999$. By comparing the model with Planck data only, they find $\epsilon=-0.43^{+0.28}_{-0.21}$\footnote{Their parameter $\xi$ corresponds to $\epsilon$ from our analysis.} and $H_0=71.7^{+2.3}_{-2.7}$ km/s/Mpc at 68\% c.l. By comparing the model with Planck+transversal BAO+$H_0$, they find $\epsilon=-0.58\pm0.11$ and $H_0=73.99\pm0.88$ km/s/Mpc at 68\% c.l. The $H_0$ values obtained from \cite{BernuiEtAl23} are compatible with our analysis. The value they have obtained, however, for the $\epsilon$ parameter are incompatible with our analysis and also indicates a decay of DM into DE, differently of what we obtained here, that is, a decay of $\Lambda$ into DM. The possible reasons for this difference is the fact that they have made an analysis with DE perturbation, which does not make sense for our model, as $\Lambda$ is expected to be smooth. They also have used different data from the current analysis, that is, the full Planck CMB power spectrum and transversal BAO data.

\subsection{\texorpdfstring{$\Lambda$}{L}CDM comparison}
{In order to compare the results we have obtained for these $\Lambda(t)$CDM model with the standard $\Lambda$CDM model, in this subsection we first perform the same analysis in the context of spatially flat $\Lambda$CDM model. The results of our analysis can be seen on Fig. \ref{lcdm1}.}

\begin{figure}[t]
    \centering
    \includegraphics[width=0.9\textwidth]{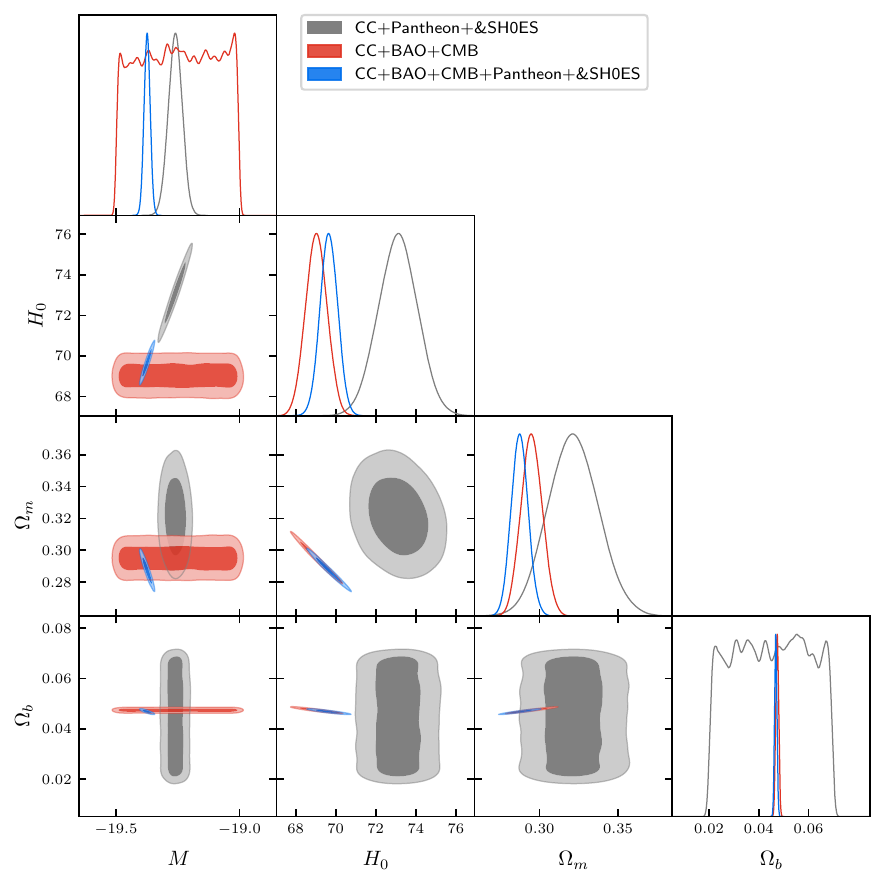}
    \caption{{Triangular plot of parameters, with data from Planck 2018 distance priors {(CMB), Pantheon+\&SH0ES, Cosmic Chronometers (CC) and transversal BAO} for flat $\Lambda$CDM model. The contours correspond to 68\% and 95\% c.l.}}\label{lcdm1}
\end{figure}

{As can be seen from Fig. \ref{lcdm1}, the constraints over $H_0$ from low redshift (mainly SH0ES) and high redshift (mainly CMB) are quite distant, with low redshift data favouring a high value for $H_0$, while high redshift data favours a low value for $H_0$, as well known in the context of $H_0$ tension. In this same figure, we also show the joint analysis, but in reality, it is meaningless, due to the high discrepancies obtained. The exact numerical results can be seen on Tab. \ref{tab3}.}


\begin{table}[ht]
    \centering
    \begin{tabular} {l | c | c | c }

Dataset & $H_0$(km/s/Mpc) & $\Omega_m$ & $\Omega_b$ \\
\hline
CC+PS & $73.08\pm 0.98  $ & $0.322\pm 0.016$ & $0.045\pm 0.015$ \\
CC+BAO+CMB & $69.02\pm 0.52$ & $0.2953\pm 0.0067$ & $0.04747\pm 0.00056$ \\
CC+PS+BAO+CMB & $69.66\pm 0.44$ & $0.2874\pm 0.0054$ & $0.04686\pm 0.00047$ \\
\hline
\end{tabular}
    \caption{{Numerical results for flat $\Lambda$CDM model. Mean values of the free parameters from the various combinations of datasets, together with the $1\sigma$ c.l. intervals.}}
    \label{tab3}
\end{table}

{As can be seen on Tab. \ref{tab3}, there was a high incompatibility between the $H_0$ constraints. In fact, the $H_0$ tension between CC+PS and CC+BAO+CMB was 3.7$\sigma$. This is lower than the usual tensions mentioned in the literature due to the fact that the transversal BAO alleviates the $H_0$ tension in the context of $\Lambda$CDM model, as already shown by \cite{BernuiEtAl23}.}

{In order to compare more directly the interacting models with $\Lambda$CDM, let us calculate two information criteria, namely, the AIC (Akaike Information Criterion) and BIC (Bayesian Information Criterion). As explained, for instance, in \cite{Akaike74,JesusEtAl17}, the AIC can be written as}
\be
\text{AIC}=-2\ln\mathcal{L}_{max}+2p
\ee
{where $\mathcal{L}_{max}$ is the maximum of the posterior distribution and $p$ is the number of parameters. While AIC is an interesting tool for comparison between models, it is taken as not penalizing too much the excess of parameters. In order to further penalize the excess of parameters, it was defined the BIC, which is considered as an approximation of the Bayesian Evidence. The BIC can be written as \cite{Schwarz78,JesusEtAl17}}
\be
\text{BIC}=-2\ln\mathcal{L}_{max}+p\ln n
\ee
{where $n$ is number of data. The results for AIC and BIC as well for $\chi^2_\nu\equiv\frac{\chi^2_{min}}{n-p}$, can be seen on Table \ref{tab4}.}

\begin{table}[ht]
    \centering
    \begin{tabular} {c | c | c | c | c | c}

Model & $\chi^2_\nu$ &  AIC & $\Delta$AIC & BIC & $\Delta$BIC\\
\hline
 I & $0.8910$ & $1514.1511$ & 0 & $1541.3548$ & 0\\
 II & $0.8853$ & $1526.6273$ & $12.4762$& $1553.8398$ & $12.485$ \\
Flat $\Lambda$CDM & $0.9012$ & $1540.1878$ & $26.0369$& $1561.9507$ & $20.5959$ \\
\hline
\end{tabular}
    \caption{{Comparison among the interacting models (I and II) and flat $\Lambda$CDM model.}}
    \label{tab4}
\end{table}

{As can be seen on Tab. \ref{tab4}, although the $\chi^2_\nu$ is similar for the three models analysed here, the $\chi^2_{min}$ for $\Lambda$CDM is too high when compared to Models I and II, in such a way that even if AIC and BIC penalize the excess of parameters of Models I and II, they still have advantage against $\Lambda$CDM. According to \cite{JesusEtAl17}, a value of $\Delta\text{AIC}>10$ and $\Delta\text{BIC}>5$ is decisive against the model in question. So, we may conclude that both Model II and flat $\Lambda$CDM model can be discarded by the current analysis.} 

\subsection{Comparison with other interacting models and analyses}

The results obtained here departs markedly from earlier running--vacuum work. With a positive interaction strength $\epsilon\simeq0.02$ {from both models} -- meaning a small energy flow from vacuum to dark matter -- the combined CC+Pantheon+\&SH0ES+BAO+Planck data yield $H_0 = 71.6\pm0.6$ km/s/Mpc and rule out $\epsilon=0$ at $\sim6\sigma$. This lifts the CMB‐based value of $H_0$ into full agreement with the local distance‐ladder measurement while leaving only a mild ($\sim2\sigma$) residual tension in $\Omega_m$. By contrast, in the running‐vacuum models of Sol\`{a} and collaborators \cite{SolaPeracaula:2017esw,Sola:2017znb}, a much smaller parameter $\nu\sim10^{-3}$ is obtained; their global fits favour $H_0\approx67$--$68\;\mathrm{km\,s^{-1}\,Mpc^{-1}}$ (close to Planck) and ease the $\sigma_8$ discrepancy.  
The early RG analyses of Shapiro et al. (2003) \cite{Shapiro:2003ui} and Espa\~na‐Bonet et al. (2004) \cite{Espana-Bonet:2003qjh} anticipated a natural percent‐level running ($\nu\sim0.02$) but lacked data to test it. {More recent RVM analyses, however, already have favoured interacting models over the $\Lambda$CDM model. In Ref. \cite{SolaPeracaula:2021gxi}, the so-called Type-I RRVM model with vacuum decay but no $G$ variation, is analysed against SNe Ia+SH0ES+BAO+$H(z)$+LSS+Planck data. From this analysis, they have found $\nu_{eff}=-0.00005^{+0.00040}_{-0.00038}$ as interacting parameter. Therefore, in this scenario, a noninteracting model with $\nu_{eff}=0$ can not be discarded. However, by adding a switch at $z_*\sim1$ for the same type of interaction, defining what they call a threshold Type-I RRVM, they find $\nu_{eff}=0.0209^{+0.0055}_{-0.0059}$, thereby discarding $\nu_{eff}=0$ at 3.5$\sigma$. More recently \cite{SolaPeracaula:2023swx}, the same models were analysed against Pantheon+\&SH0ES+BAO+$H(z)$+LSS+Planck data, and they have found $\nu_{eff}=-0.00037\pm0.00029$ for the Type-I RRVM and $\nu_{eff}=0.0197\pm0.0055$ for the threshold Type-I RVM model. Again, $\nu_{eff}$ could not be discarded in the Type-I RRVM scenario, while in the Type-I RRVM scenario, the $\Lambda$CDM model could be discarded at $3.6\sigma$ c.l.}

By comparing our analysis with these recent works, we may conclude that the models studied here are simpler and are better favoured against $\Lambda$CDM, concerning the $H_0$ tension. On the other hand, these recent RVM studies are more complete, in the sense that they use the full Planck likelihood and LSS data, and also in the sense that they attack both $H_0$ and $\sigma_8$ tensions.

More recently, while the present work was in the submission stage and we were improving and extending the analysis, a new paper has been published \cite{YangEtAl25}, based on our original proposal of studying these models. In \cite{YangEtAl25}, they have studied the Models I and II. However, their analysis is slightly different concerning the CC dataset, where we have used 32 $H(z)$ data from \cite{MorescoEtAl22} and they have used just 31 $H(z)$ data from various references. Concerning the CMB distance prior, we are unable to compare their analysis with the present analysis, as they do not claim which distance prior they have chosen to use from Ref. \cite{ChenEtAl19}. In the present work, we have chosen to work only with the flat $\Lambda$CDM distance prior. However, we do not expect to find an appreciable difference for changing the distance prior to $w$CDM for instance, as these priors are quite similar. Despite these slight differences, when we compare the SNe Ia data we have used in the present work with the one by \cite{YangEtAl25}, the difference is huge: They have used the Pantheon compilation \cite{Pantheon}, which contains only 1048 SNe Ia data and do not constrain $H_0$ alone, while we have used Pantheon+\&SH0ES \cite{pantheon+}, which consists of 1701 light curves of 1550 distinct SNe Ia, and strongly constrains $H_0$. In \cite{YangEtAl25}, they have found a 2.8$\sigma$ evidence for an interaction in the context of Model I and 2.6$\sigma$ evidence for interaction in the context of Model II. Since we have found at least a $\sim5.8\sigma$ evidence for interaction in the context of these models, we attribute this difference to the fact that we have used Pantheon+\&SH0ES, which strongly constrains $H_0$, while \cite{YangEtAl25} have chosen not to use $H_0$ constraints from SH0ES in their analysis. It is also important to note that \cite{YangEtAl25} have used the DESI DR1 BAO data \cite{DESIDR1}, while we have chosen to use the transversal BAO data \cite{NunesEtAl20}. It is important to mention that while DESI uses 3D BAO, the transversal BAO data consists of 2D BAO. In recent works \cite{ValentSola24,ValentSola24ApJ}, both analyses have been made in the context of dynamical dark energy models, and a higher evidence were found for dynamical dark energy in the case of 2D BAO. The fact that we have chosen to use 2D BAO in the present work may also contribute to a higher evidence of dark matter-quantum vacuum interaction. This has also appeared on Ref. \cite{BernuiEtAl23}, where they have shown that the $H_0$ tension disappears in the context of Model I by using 2D BAO data, while it is only alleviated when using 3D BAO. We have to remember, however, that the 2D BAO data is weakly dependent of cosmological models, as mentioned by \cite{NunesEtAl20}, which makes it more suitable to analyse alternative models as Models I and II.


\section{\label{sec: conclusion}Conclusion}
We have analysed the Hubble tension within the context of dark energy models with a time-varying cosmological constant, specifically vacuum decay models. The results obtained in this work show that vacuum decay models $  \Lambda(t)$CDM, with interaction between dark matter and the  cosmological parameter, are promising in addressing the Hubble $H_0$ tension. Our results show that vacuum decay models, where the cosmological term $\Lambda$ varies over time, can reconcile the $H_0$ estimates, providing higher values in alignment with the {low-redshift, Pantheon+\&SH0ES data. It is clear from our analysis that the decay parameter $\epsilon$ allows $H_0$ to reach higher values for the high-redshift constraints, thereby reconciling high-redshift constraints with the local value for $H_0$. As a result of the interaction, however, for both models, a higher value for $\Omega_m$ is predicted from low-redshift data and a lower value for $\Omega_m$ is predicted from high-redshift data. This tension is around $2\sigma$, representing a transfer and an alleviation of the $H_0$ tension. Due to the positive correlation between $\epsilon$ and $H_0$, a higher value for $H_0$ results in a positive value for $\epsilon$, according to our analysis.}
From a thermodynamic point of view it is well known that this is the preferred direction of decay, as it does not require a non-vanishing chemical potential in the dark sector \cite{PereiraJesus09}. This suggests that the energy flow from vacuum to matter {can be} compatible with a higher $H_0$. {Although still showing some tension concerning the matter density parameter, our study shows that interacting models can be a promising direction to solve the $H_0$ tension.} {Therefore, it would be interesting, in the future, that a full CMB and LSS analyses would be performed in the context of these models, thereby providing a more definitive conclusion about the ability of $\Lambda(t)$CDM of solving or alleviating not only the $H_0$ tension, but also the $\sigma_8$ tension.}

\begin{acknowledgments}
This study was financed in part by the Coordena\c{c}\~ao de Aperfei\c{c}oamento de Pessoal de N\'ivel Superior - Brasil (CAPES) - Finance Code 001. JFJ acknowledges financial support from  {Conselho Nacional de Desenvolvimento Cient\'ifico e Tecnol\'ogico} (CNPq) (No. 314028/2023-4). SHP acknowledges financial support from  {Conselho Nacional de Desenvolvimento Cient\'ifico e Tecnol\'ogico} (CNPq)  (No. 308469/2021 and 301775/2025-7). 
\end{acknowledgments}


\begin{thebibliography}{30}

\bibitem{Perivolaropoulos2022}
L.~Perivolaropoulos and F.~Skara,
New Astron. Rev. \textbf{95} (2022), 101659
[arXiv:2105.05208 [astro-ph.CO]].

\bibitem{Weinberg:1988cp}
S.~Weinberg,
Rev. Mod. Phys. \textbf{61} (1989), 1-23

\bibitem{Velten:2014nra}
H.~E.~S.~Velten, R.~F.~vom Marttens and W.~Zimdahl,
Eur. Phys. J. C \textbf{74} (2014) no.11, 3160
[arXiv:1410.2509 [astro-ph.CO]].

\bibitem{H0TensionReview}
E.~Di Valentino, O.~Mena, S.~Pan, L.~Visinelli, W.~Yang, A.~Melchiorri, D.~F.~Mota, A.~G.~Riess and J.~Silk,
Class. Quant. Grav. \textbf{38} (2021) no.15, 153001
[arXiv:2103.01183 [astro-ph.CO]].

\bibitem{Riess:2009pu}
A.~G.~Riess, L.~Macri, S.~Casertano, M.~Sosey, H.~Lampeitl, H.~C.~Ferguson, A.~V.~Filippenko, S.~W.~Jha, W.~Li and R.~Chornock, \textit{et al.}
Astrophys. J. \textbf{699} (2009), 539-563
[arXiv:0905.0695 [astro-ph.CO]].

\bibitem{BreuvalEtAl24}
L.~Breuval, A.~G.~Riess, S.~Casertano, W.~Yuan, L.~M.~Macri, M.~Romaniello, Y.~S.~Murakami, D.~Scolnic, G.~S.~Anand and I.~Soszy\'nski,
Astrophys. J. \textbf{973} (2024) no.1, 30
[arXiv:2404.08038 [astro-ph.CO]].


\bibitem{Planck18}
N.~Aghanim \textit{et al.} [Planck],
Astron. Astrophys. \textbf{641} (2020), A6
[erratum: Astron. Astrophys. \textbf{652} (2021), C4]
[arXiv:1807.06209 [astro-ph.CO]].




\bibitem{Ahmet2018}
A.~M.~\"Ozta\c{s},
Mon. Not. Roy. Astron. Soc. \textbf{481} (2018) no.2, 2228-2234


\bibitem{Vishwakarma2001}
R.~G.~Vishwakarma,
Class. Quant. Grav. \textbf{18} (2001), 1159-1172
[arXiv:astro-ph/0012492 [astro-ph]].


\bibitem{Overduin1998}
J.~M.~Overduin and F.~I.~Cooperstock,
Phys. Rev. D \textbf{58} (1998), 043506
[arXiv:astro-ph/9805260 [astro-ph]].


\bibitem{Azri2017}
H.~Azri and A.~Bounames,
Int. J. Mod. Phys. D \textbf{26} (2017) no.7, 1750060
[arXiv:1412.7567 [gr-qc]].


\bibitem{Azri2012}
H.~Azri and A.~Bounames,
Gen. Rel. Grav. \textbf{44} (2012), 2547-2561
[arXiv:1007.1948 [gr-qc]].


\bibitem{Szydlowski2015}
M.~Szyd\l{}owski,
Phys. Rev. D \textbf{91} (2015) no.12, 123538
[arXiv:1502.04737 [astro-ph.CO]].


\bibitem{Bruni2022}
M.~Bruni, R.~Maier and D.~Wands,
Phys. Rev. D \textbf{105} (2022) no.6, 063532
[arXiv:2111.01765 [gr-qc]].


\bibitem{Papagian2020}
G.~Papagiannopoulos, P.~Tsiapi, S.~Basilakos and A.~Paliathanasis,
Eur. Phys. J. C \textbf{80} (2020) no.1, 55
[arXiv:1911.12431 [gr-qc]].


\bibitem{Benetti2019}
M.~Benetti, W.~Miranda, H.~A.~Borges, C.~Pigozzo, S.~Carneiro and J.~S.~Alcaniz,
JCAP \textbf{12} (2019), 023
[arXiv:1908.07213 [astro-ph.CO]].


\bibitem{Benetti2021}
M.~Benetti, H.~Borges, C.~Pigozzo, S.~Carneiro and J.~Alcaniz,
JCAP \textbf{08} (2021), 014
[arXiv:2102.10123 [astro-ph.CO]].

\bibitem{Macedo:2023zrd}
H.~A.~P.~Macedo, L.~S.~Brito, J.~F.~Jesus and M.~E.~S.~Alves,
Eur. Phys. J. C \textbf{83} (2023) no.12, 1144,
[arXiv:2305.18591 [astro-ph.CO]].





\bibitem{Moreno-Pulido:2020anb}
C.~Moreno-Pulido and J.~Sola,
Eur. Phys. J. C \textbf{80} (2020) no.8, 692,
[arXiv:2005.03164 [gr-qc]].

\bibitem{Moreno-Pulido:2022phq}
C.~Moreno-Pulido and J.~Sola Peracaula,
Eur. Phys. J. C \textbf{82} (2022) no.6, 551,
[arXiv:2201.05827 [gr-qc]].

\bibitem{Moreno-Pulido:2023ryo}
C.~Moreno-Pulido, J.~Sola Peracaula and S.~Cheraghchi,
Eur. Phys. J. C \textbf{83} (2023) no.7, 637,
[arXiv:2301.05205 [gr-qc]].

\bibitem{Sola:2013gha}
J.~Sola,
J. Phys. Conf. Ser. \textbf{453} (2013), 012015,
[arXiv:1306.1527 [gr-qc]].

\bibitem{SolaPeracaula:2022hpd}
J.~Sola Peracaula,
Phil. Trans. Roy. Soc. Lond. A \textbf{380} (2022), 20210182,
[arXiv:2203.13757 [gr-qc]].



\bibitem{SolaPeracaula:2021gxi}
J.~Sol\`a Peracaula, A.~G\'omez-Valent, J.~de Cruz Perez and C.~Moreno-Pulido,
EPL \textbf{134} (2021) no.1, 19001
[arXiv:2102.12758 [astro-ph.CO]].

\bibitem{SolaPeracaula:2017esw}
J.~Sol\`a Peracaula, J.~de Cruz P\'erez and A.~Gomez-Valent,
Mon. Not. Roy. Astron. Soc. \textbf{478} (2018) no.4, 4357-4373,
[arXiv:1703.08218 [astro-ph.CO]].


\bibitem{Sola:2017znb}
J.~Sol\`a, A.~G\'omez-Valent and J.~de Cruz P\'erez,
Phys. Lett. B \textbf{774} (2017), 317-324,
[arXiv:1705.06723 [astro-ph.CO]].

\bibitem{SolaPeracaula:2023swx}
J.~Sola Peracaula, A.~Gomez-Valent, J.~de Cruz Perez and C.~Moreno-Pulido,
Universe \textbf{9} (2023) no.6, 262
[arXiv:2304.11157 [astro-ph.CO]].

\bibitem{Sola:2016jky}
J.~Sol\`a, A.~G\'omez-Valent and J.~de Cruz P\'erez,
Astrophys. J. \textbf{836} (2017) no.1, 43
[arXiv:1602.02103 [astro-ph.CO]].

\bibitem{Sola:2015wwa}
J.~Sola, A.~Gomez-Valent and J.~de Cruz P\'erez,
Astrophys. J. Lett. \textbf{811} (2015), L14
[arXiv:1506.05793 [gr-qc]].

\bibitem{Espana-Bonet:2003qjh}
C.~Espana-Bonet, P.~Ruiz-Lapuente, I.~L.~Shapiro and J.~Sola,
JCAP \textbf{02} (2004), 006,
[arXiv:hep-ph/0311171 [hep-ph]].

\bibitem{Shapiro:2003ui}
I.~L.~Shapiro, J.~Sola, C.~Espana-Bonet and P.~Ruiz-Lapuente,
Phys. Lett. B \textbf{574} (2003), 149-155,
[arXiv:astro-ph/0303306 [astro-ph]].



\bibitem{PereiraJesus09}
S.~H.~Pereira and J.~F.~Jesus,
Phys. Rev. D \textbf{79} (2009), 043517
[arXiv:0811.0099 [astro-ph]].

\bibitem{PavonZimdahl05}
D.~Pavon and W.~Zimdahl,
Phys. Lett. B \textbf{628} (2005), 206-210
[arXiv:gr-qc/0505020 [gr-qc]].

\bibitem{PavonWang09}
D.~Pavon and B.~Wang,
Gen. Rel. Grav. \textbf{41} (2009), 1-5
[arXiv:0712.0565 [gr-qc]].

\bibitem{Chen1990}
W.~Chen and Y.~S.~Wu,
Phys. Rev. D \textbf{41} (1990), 695-698
[erratum: Phys. Rev. D \textbf{45} (1992), 4728]

\bibitem{JohnJoseph99}
M.~V.~John and K.~B.~Joseph,
Phys. Rev. D \textbf{61} (2000), 087304
[arXiv:gr-qc/9912069 [gr-qc]].

\bibitem{Rajeev83}
S.~G.~Rajeev,
Phys. Lett. B \textbf{125} (1983), 144-146

\bibitem{pantheon+}
D.~Brout, D.~Scolnic, B.~Popovic, A.~G.~Riess, J.~Zuntz, R.~Kessler, A.~Carr, T.~M.~Davis, S.~Hinton and D.~Jones, \textit{et al.}
Astrophys. J. \textbf{938} (2022) no.2, 110
[arXiv:2202.04077 [astro-ph.CO]].

\bibitem{MorescoEtAl22}
M.~Moresco, L.~Amati, L.~Amendola, S.~Birrer, J.~P.~Blakeslee, M.~Cantiello, A.~Cimatti, J.~Darling, M.~Della Valle and M.~Fishbach, \textit{et al.}
Living Rev. Rel. \textbf{25} (2022) no.1, 6
[arXiv:2201.07241 [astro-ph.CO]].

\bibitem{ChenEtAl19}
L.~Chen, Q.~G.~Huang and K.~Wang,
JCAP \textbf{02} (2019), 028
[arXiv:1808.05724 [astro-ph.CO]].

\bibitem{NunesEtAl20}
R.~C.~Nunes, S.~K.~Yadav, J.~F.~Jesus and A.~Bernui,
Mon. Not. Roy. Astron. Soc. \textbf{497} (2020) no.2, 2133-2141
[arXiv:2002.09293 [astro-ph.CO]].

\bibitem{emcee}
D.~Foreman-Mackey, D.~W.~Hogg, D.~Lang and J.~Goodman,
Publ. Astron. Soc. Pac. \textbf{125} (2013), 306-312
[arXiv:1202.3665 [astro-ph.IM]].

\bibitem{getdist}
A.~Lewis,
[arXiv:1910.13970 [astro-ph.IM]].

\bibitem{BernuiEtAl23}
A.~Bernui, E.~Di Valentino, W.~Giar\`e, S.~Kumar and R.~C.~Nunes,
Phys. Rev. D \textbf{107} (2023) no.10, 103531
[arXiv:2301.06097 [astro-ph.CO]].

\bibitem{JesusEtAl17}
J.~F.~Jesus, R.~Valentim and F.~Andrade-Oliveira,
JCAP \textbf{09} (2017), 030
[arXiv:1612.04077 [astro-ph.CO]].

\bibitem{Akaike74}
H.~Akaike,
IEEE Trans. Automatic Control \textbf{19} (1974) no.6, 716-723

\bibitem{Schwarz78}
G.~Schwarz,
Annals Statist. \textbf{6} (1978), 461-464

\bibitem{YangEtAl25}
Y.~Yang, Y.~Wang and X.~Dai,
Eur. Phys. J. C \textbf{85} (2025) no.3, 224
[arXiv:2502.17792 [astro-ph.CO]].

\bibitem{Pantheon}
D.~M.~Scolnic \textit{et al.} [Pan-STARRS1],
Astrophys. J. \textbf{859} (2018) no.2, 101
[arXiv:1710.00845 [astro-ph.CO]].

\bibitem{DESIDR1}
A.~G.~Adame \textit{et al.} [DESI],
JCAP \textbf{02} (2025), 021
doi:10.1088/1475-7516/2025/02/021
[arXiv:2404.03002 [astro-ph.CO]].

\bibitem{ValentSola24}
A.~G{\'o}mez-Valent and J.~Sol{\`a} Peracaula,
Phys. Lett. B \textbf{864} (2025), 139391
[arXiv:2412.15124 [astro-ph.CO]].

\bibitem{ValentSola24ApJ}
A.~Gomez-Valent and J.~Sol{\`a} Peracaula,
Astrophys. J. \textbf{975} (2024) no.1, 64
[arXiv:2404.18845 [astro-ph.CO]].

\end{thebibliography}


\end{document}